# Using electropolymerization based doping for electro-addressable functionalization of a multi-electrode array probe for nucleic acid detection

**Mustafa Şen**[1,2]
[1] Biomedical Engineering Department, Izmir Katip Celebi University, Izmir, Turkey
[2] Biomedical Technologies Graduate Program, Izmir Katip Celebi University, Izmir, Turkey

E-mail: mustafa.sen@ikc.edu.tr

Here, we report a facile method for electro-addressable functionalization of a probe comprising of closely spaced three individually addressable carbon fiber electrodes for detection of nucleic acids. First, a multi electrode array probe comprising three adjacent carbon fiber electrodes was fabricated through pulling a three-barrel glass capillary with a single carbon fiber in each barrel using a micropuller. Second, electropolymerization based doping was used for electro-addressable functionalization of the multi-electrode array probe. To demonstrate that the current strategy works, anti-miR-34a was electrografted on only one of three electrodes by electropolymerization of pyrrole on the specific electrode. A second electrode was coated only with polypyrrole (PPy) and the third was left unmodified. Electrochemical impedance spectroscopy (EIS) was used for analysis and the results clearly showed charge transfer resistance of the PPy + anti-miR-34a modified electrode increased significantly after hybridization, while charge transfer resistance of the other two electrodes remained almost constant. The results demonstrate that the present strategy has great potential for constructing multiplex nucleic acid micro/nano biosensors for local and in situ detection of multiple nucleic acid molecules such as miRNAs at a time.

The use of microelectrodes in construction of microbiosensors has many advantages such as the ability to achieve high temporal and spatial resolution, local detection and detection in small volumes[1]. Increasing the number of electrodes enables detection of multiple analytes or simultaneous and high-throughput detection of same molecules at multiple points. Even though large number of individually addressable electrodes can be effectively assembled in a small area[2-5], electrochemical detection methods have a relatively limited ability to detect multiple analytes, particularly when compared to microdialysis probes and other analytic techniques based on chromatography or electrophoretic separation [6]. There are several reports where multi-electrode arrays were assembled on a substrate to form a biosensor chip to perform multiple analysis. For example, Zhu et al. used a multi-point addressable electrochemical device where row and column electrodes were orthogonally arranged to form a 4 x 4 array for amperometric detection of DNA hybridization.[7] Because of the large distance between electrodes, they were able to modify electrodes with different probes using a pipette. The device had a detection limit of 30 nM. In another study, Erdem et al. used a multi-channel screen printed array electrode for label-free voltammetric detection of microRNAs.[8] Electrode array comprising 16 relatively large carbon electrodes enabled 16 consecutive analysis with a detection limit of 4.3 pmole in 3 μL sample. The treatment of the electrodes with different samples was also done by pipetting due to large distance between the electrodes. A microelectrode array comprising closely spaced electrodes with different labels enables simultaneous and local detection of different analytes in small volumes, but labeling closely spaced electrodes with different bio-recognition molecules is technically challenging as pipetting is not an option. In a recent study, Ozsoz at al. used polypyrrole (PPy) for the doping of probe miRNA on a large pencil graphite electrode to detect miR-21[9]. PPy is one of the most widely used conducting polymer for biosensor design due to its biocompatibility, high electrical conductivity, wide pH range, fast electrochemical reaction, high surface energy and low cost. Basically, in the study, the miRNA probe was mixed with the pyrrole solution and by electropolymerization of pyrrole, the miRNA probes were electrografted on the electrode surface. The biosensor showed good selectivity and had a detection limit of 0.17 nM. Since pyrrole is electropolymerized only on the electrode through which the electropolymerization potential is applied, we used it for electro-addressable functionalization of a closely spaced multi carbon fiber electrode array micro-probe with a pitch of less than 5 μm for the detection of miR-34a. Carbon fiber based microelectrodes are commonly used for local and highly sensitive detection of neurotransmitters such as dopamine or recording of neuronal action potentials known as spikes, enabling electrochemical monitoring of neurochemical activity of brain [10-13]. In addition to being used in such measurements, CFEs have also been used for detection of biologically important molecules with great sensitivity and selectivity by simply modifying the surface of the electrode with a bio-recognition molecule [14-17]. To the best our knowledge this is the first study to use a probe-type microelectrode array for nucleic acid detection. The probe was used in self-referencing mode[18], for detection of miR-34a, where one of three electrodes was modified with anti-miRNA34a using PPy. A second electrode was coated with PPy and the third was left unmodified. The results clearly demonstrate that electropolymerization based doping works for electro-addressable functionalization of a micro-electrode array probe and has a great potential for constructing micro-biosensors for detection of multiple miRNAs at a time

For fabrication of the multi-electrode array probe, three copper wires, each connected to a single carbon fiber, were prepared. Each wire was prepared as follows; a single carbon fiber with a diameter of 5-6 μm was fixed to an insulated copper wire. To make the connection, one end of the copper wire was dipped into Ag paste that was then used for attaching a single carbon fiber. The Ag paste was cured at 180 °C for 30 min to make a rigid connection between the carbon fiber and the copper wire. Three copper wires, each fixed with a carbon fiber, were individually inserted into a three-barrel capillary. To fabricate a multi-electrode array probe with a thin and parallel run to the very end of the tip, the three-barrel capillary was pulled (PG10165-4, World Precision Instrument, USA) with a micropuller (PC-10, Narishige, Japan) using the one-stage pull option to seal the carbon fibers in glass (Figure 1A)[1,19-21]. The heating parameter that yielded a multi-electrode array probe with the desired characteristics was 68 °C. In the following step, the excessive unsealed carbon fibers at the tip of the probe were removed with a scissor. To fill any possible gap between the carbon-fibers and the sealed glass, the tip of the probe was dipped into resin which was then cured at 80 °C for 40 min. To fabricate micro-disk electrodes, the tip of the probe was ground using a microgrinder (EG-401, Narishige, Japan) (Figure 1B).

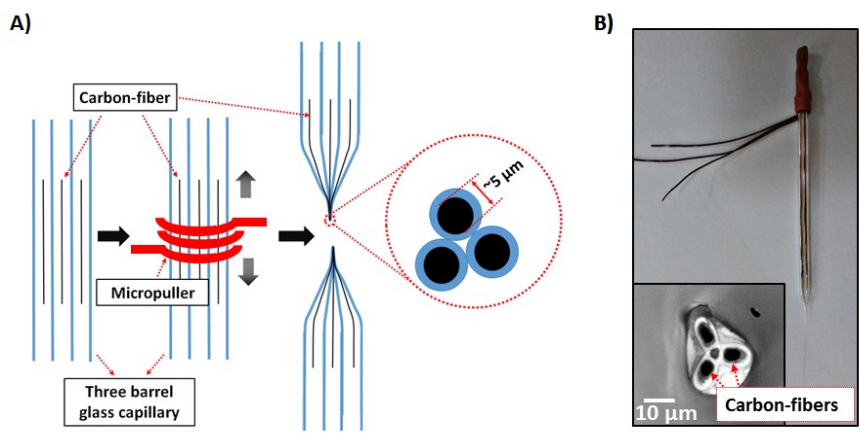

**Figure 1.** (A) Schematic illustration showing the fabrication method for multi electrode array probe. First, three carbon fibers were individually inserted into a three-barrel glass capillary and then the capillary was pulled with a micropuller. (B) An optical image of the probe along with an image of the tip showing the three electrodes.

The electrochemical behavior of the carbon fiber electrodes was analyzed to assess the usability of the probe. First, cyclic voltammograms (CVs) of each electrode were obtained by sweeping the potential from 0 to +0.6 V vs. Ag/AgCl at a scan rate of 50 mV/sec in 1 mM ferrocenemethanol (FcCH$_2$OH + PBS). Second, a simultaneous CV of all electrodes was obtained under the same conditions. All three electrodes showed similar behavior; the maximum current was measured to be ~1.24 nA (Figure 2A), quite close to the simulated current value of ~1.16 nA (Figure 2B) found using COMSOL Multiphysics. We believe the difference between experimental and simulated values is normal considering the likely non-ideal surface conditions of the tips of the probe. When all three electrodes were connected, the maximum current measured was ~2.71 nA (Figure 2A), approximately 2.2 times of the current obtained from a single electrode and quite close to the simulated value of ~2.74 nA (Figure 2B). These results clearly indicate that all electrodes are working properly and individually addressable.

After confirming the integrity of the electrodes, the multi-electrode array probe was modified for miR-34a detection using electropolymerization based doping (Figure 3A). The miR-34a, a non-coding miRNA, is a direct transcriptional target of the onco-suppressor p53, which plays a critical role in genomic stability, apoptosis, and inhibition of angiogenesis [22,23]. To show each electrode could be used for a different purpose, only one of three electrodes was modified with anti-miR-34a using PPy; 5 ppm anti-miR-

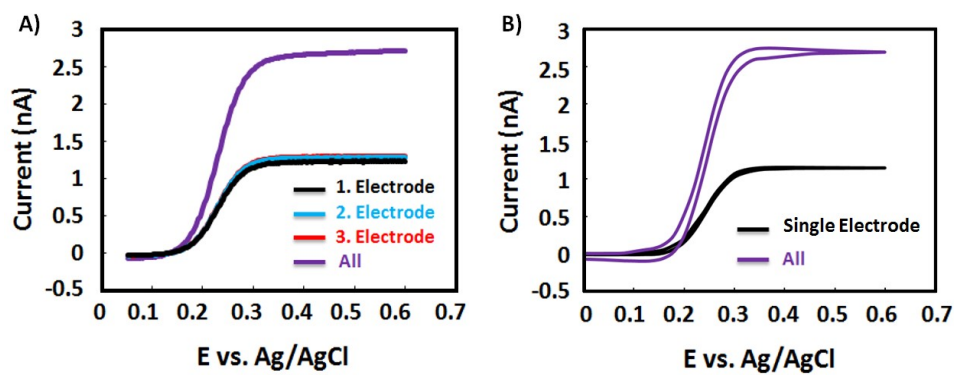

**Figure 2.** Experimental (A) and simulated (B) cyclic voltammograms of individual carbon-fiber electrodes in the same multi electrode array probe. All electrodes numbered form 1 to 3 demonstrated identical electrochemical behavior. CV when all connected is also shown as "All" in both graphs. Electrode numbers correspond to the electrode numbers given in the illustration of the next figure (Figure 3).

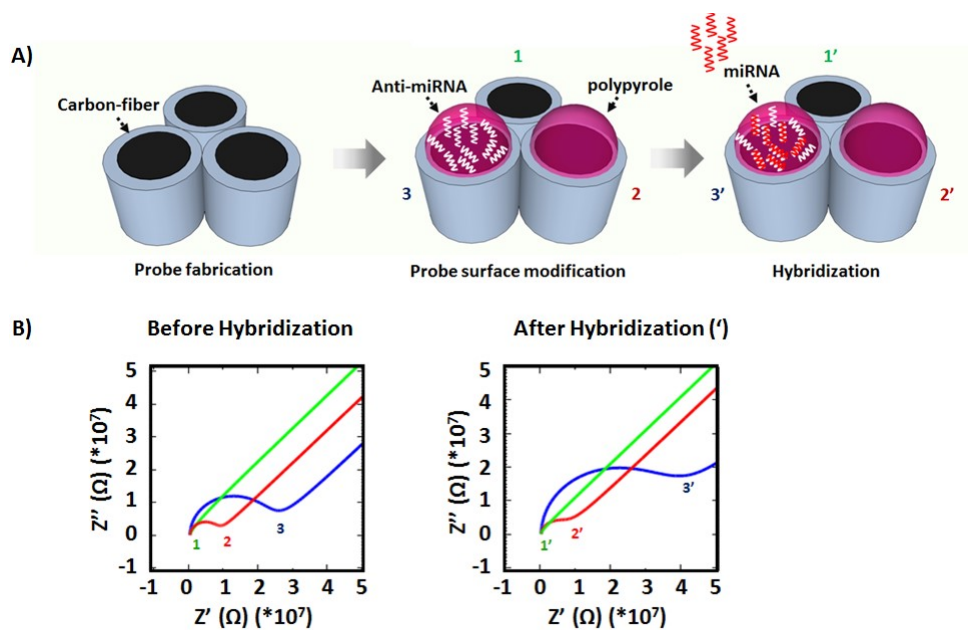

**Figure 3.** (A) Modification of the multi electrode array probe for miR-34a detection. The probe was used in "self-referencing mode" for detection of miR-34a, in which case the first electrode (1) was left unmodified while second (2) and third (3) carbon-fiber electrodes were modified with PPy and anti-miRNA34a + PPy, respectively. (B) Electrochemical impedance spectroscopy (EIS) results before (1, 2 and 3) and after (1', 2' and 3') hybridization. Numbers in the graphs represent the numbers assigned for the electrodes in the illustration.

34a was mixed with PPy and the probe, along with an Ag/AgCl reference electrode, was immersed into the mixture. PPy with the anti-miR-34a was electrochemically deposited by sweeping the potential from 0 to +0.8 V vs. Ag/AgCl four times at a scan rate of 50 mV/sec. Next, a second electrode was modified with PPy in the same way but without anti-miR-34a. The last electrode was left unmodified. Electrochemical impedance spectroscopy (EIS) was then employed for impedimetric analysis of the electrodes for confirmation of the modification and comparison after hybridization. The experiments were performed in a 5 mM $[Fe(CN)_6]^{3-/4-}$ and 0.1 M KCl solution using a AUTOLAB PGSTAT204 compact and modular potentiostat/galvanostat. The impedance spectra were recorded in a frequency range of $10^1$ to $10^5$ Hz, which was divided into 98 logarithmically equidistant points. A DC potential of +0.23 V vs. Ag/AgCl along with a sinusoidal AC signal of 10 mV was applied throughout all EIS measurements, which were performed in a faraday cage. The data were fitted to an equivalent circuit and the diameter of first semicircle of the Nyquist plot, corresponding to the charge transfer resistance ($R_{ct}$) of the electrode surface, was calculated by the fit and simulation option in the AUTOLAB 302 Nova 1.10.3 software. According to the results, the impedance response of the unmodified electrode exhibited a straight line with almost no semicircular shape. The semicircle shape of the PPy modified electrode ($R_{ct}$: 7.68 MΩ) plot was lower than that of anti-miR-34a + PPy modified electrode ($R_{ct}$: 22.5 MΩ) because anti-miR-34a increases the resistance. In other words, these results clearly indicate that anti-miR-34a was successfully deposited electrochemically to one of the electrodes.

The modified probe was immersed into a 50 mM PBS at pH 7.4 solution containing 10 ppm of miR-34a at room temperature for 1 h to allow miR-34a hybridize with anti-miR-34a on the electrode surface. Once this step was complete, the probe was washed in 50 mM PBS at pH 7.4 for 10 sec. Impedimetric analysis of the electrodes was performed in a 5 mM $K_3Fe[(CN)_6]$ and 0.1 M KCl solution to verify miR-34a detection. Figure 3B shows that the data of the unmodified (almost no semicircular shape) and PPy ($R_{ct}$: 6.63 MΩ or 6.29 Ω.cm$^2$) modified electrodes remained the same whereas the semicircular shape of the anti-miR34a modified electrode increased substantially after hybridization (the $R_{ct}$ value increased from 22.5 (21.37 Ω.cm$^2$) to 34.5 MΩ (32.76 Ω.cm$^2$)). A significant change was observed in the phase angle Bode plots of the CFE after surface modification. When the phase angle Bode plots of PPy + anti-miR-34a modified electrode before and after hybridization with miR-34a were compared, the maximum change in phase was found at low frequencies, which suggests that the capacitive behavior of the biosensor is dominant at low frequencies. The change also indicates the presence of capacitive binding as the phase angle after hybridization increased[24]. In other words, miR-34a was successfully detected using a modified electrode in a multi-electrode array probe. In order to show the bio-selectivity of the biosensor, a probe whose surface was modified with PPy + anti-miR-34a was dipped in a solution containing 10 ppm non-complementary miR-21 for 1h at room temperature and no significant difference was found in charge transfer resistance (ΔRct) even though there were 7 matching bases. To confirm repeatability, the experiment was repeated three times and average ΔRct values obtained for PPy and PPy + anti-miR-

34a modified electrodes after hybridization. Using the probe in this manner enables self-referencing, since all three electrodes share the same conditions. Self-referencing helps remove interference of the analytical signal and makes detection more accurate.

In conclusion, we used electropolymerization based doping for electro-addressable functionalization of a multi-electrode array probe with three electrodes for detection of miR-34a, a non-coding miRNA that has a significant impact on apoptosis. The fabrication method used in this study yielded a multi-electrode array probe comprising of three electrodes and all electrodes demonstrated identical electrochemical behavior. The probe was used in self-referencing mode for detection of miR-34a with one of the electrodes modified with PPy + anti-miR-34a. Instead of using separate electrodes for both the test and control groups as used in a conventional analysis, a single probe with three electrodes, each assigned for a different measurement, was used in this mode. Because all three electrodes share the same microenvironment, all measurements were completed with a single probe which made the analysis comparatively easy and accurate. Due its size, the probe can be easily used in very small volumes such as single cell analysis as well as in situ detection. Although the multi-electrode array probe fabricated had three electrodes, it is possible to fabricate micro or even nano multi-electrode array probes with more electrodes using capillaries with more barrels and electropolymerization based doping can be easily employed for electro-addressable functionalization of electrodes to detect a large number of different miRNAs at a time[25,26]. In brief, miniaturization of miRNA sensor is meaningful for constructing multiplex miRNA sensor, *in vivo* monitoring of local miRNA and reducing sample volume.

This research was partly supported by Scientific Research and Project Coordinatorship of Izmir Katip Celebi University (No. 2015-GAP-MÜMF-0013) and the Scientific Council of Turkey (TUBITAK) (No. 115C093).